\documentclass[reprint, NumberedRefs]{JASA-modif-postpr}

\usepackage[protrusion=true,
            expansion=true,
            final,
            babel
                ]{microtype}
\usepackage{url}
\usepackage{setspace}
\DeclareMathOperator*{\argmax}{argmax}
\DeclareMathOperator*{\jinc}{jinc}

\begin{document}
\def\arraystretch{1.0}\ifgrouped\clo@groupedaddress\fi

\title[Arnestad \emph{et al.} \quad \emph{E-print typeset by the authors}]{Worst-case analysis of array beampatterns using interval arithmetic}
\author{Håvard Kjellmo Arnestad}
\email{haavaarn@ifi.uio.no}
\affiliation{Department of Informatics, University of Oslo, 0316 Oslo, Norway}

\author{G{\'a}bor Ger{\'e}b}
\affiliation{Department of Informatics, University of Oslo, 0316 Oslo, Norway}

\author{Tor Inge Birkenes Lønmo}
\affiliation{Kongsberg Discovery AS, 3183 Horten, Norway} 

\author{\newline{}Jan Egil Kirkebø}
\affiliation{InPhase Solutions AS, 0738 Trondheim, Norway}
 
\author{Andreas Austeng}	
\affiliation{Department of Informatics, University of Oslo, 0316 Oslo, Norway}

\author{Sven Peter Näsholm}	
\affiliation{Department of Informatics, University of Oslo, 0316 Oslo, Norway}


\date{19 June 2023} 
\DOInumber{FJDSKLJFAKL}


\begin{abstract}
Over the past decade, interval arithmetic (IA) has been utilized to determine tolerance bounds of phased array beampatterns. IA only requires that the errors of the array elements are bounded, and can provide reliable beampattern bounds even when a statistical model is missing. However, previous research has not explored the use of IA to find the error realizations responsible for achieving specific bounds. In this study, the capabilities of IA are extended by introducing the concept of ``backtracking'', which provides a direct way of addressing how specific bounds can be attained. Backtracking allows for the recovery of both the specific error realization and the corresponding beampattern, enabling the study and verification of which errors result in the worst-case array performance in terms of the peak sidelobe level. Moreover, IA is made applicable to a wider range of arrays by adding support for arbitrary array geometries with directive elements and mutual coupling, in addition to element amplitude, phase, and positioning errors. Lastly, a simple formula for approximate bounds of uniformly bounded errors is derived and numerically verified. This formula gives insights into how array size and apodization cannot reduce the worst-case peak sidelobe level beyond a certain limit.


\smallskip

\noindent\fbox{\begin{minipage}{.99\textwidth}{\flushleft\sffamily 
This article may be downloaded for personal use only. Any other use requires author and AIP Publishing prior permission. This article appears in The Journal of the Acoustical Society of America
and may be found at \url{https://doi.org/10.1121/10.0019715}.
The current e-print was typeset by the authors and can differ in, e.g., pagination, reference numbering, and typographic detail. 
Copyright 2023 The Authors. This article is distributed under a Creative Commons Attribution (CC BY) License.
}
\end{minipage}}
\end{abstract}


\maketitle





\section{\label{sec:1} Introduction}

The push for reliable, high-performance sonar array systems raises a need for analysis methods that can account for various tolerances in manufacturing and data processing. These tolerances relate to deviations from the sonar specifications such as: manufacturing imperfections, calibration tolerances, electronic processing limitations, varying environmental factors, and component wear and tear. Typically, such deviations manifest themselves as errors in the transducer element amplitude, phase response, or element mutual coupling (also referred to as cross-talk).

The beampattern of an ideally calibrated array is a function of the array geometry and electronic processing. The beampattern relates to the array's lateral resolution given by its mainlobe width. The contrast depends on the sidelobe levels, and a low contrast (i.e., high sidelobe levels) may impede target detection. This article deals with arrays subject to bounded errors and their associated beampattern bounds. These bounds determine the limits within which all possible beampattern realizations exist, and they must be constrained if one assumes that the errors are bounded. This problem is tackled using the mathematical technique of interval arithmetic (IA). The theory mainly applies to systems of small relative bandwidth, with sonars in mind. 

An analysis of array errors may be carried out statistically, as has been done since the early phased-array systems\citep{ruze_effect_1952}, and in acoustic arrays\cite{quazi_array_1982, cox_effects_2015}. The common assumption is that the relevant errors are independent and identically distributed across elements, typically Gaussian, from which one can derive that the beam\-pattern magnitude follows a Rician distribution, or more generally a Beckmann distribution\citep{van_den_biggelaar_improved_2018}. A key finding from the statistical analysis is the expression for the \emph{expected beampattern}\citep{van_trees_optimum_2004}, where a constant term due to errors or failed elements\citep{mailloux_phased_2005} may swamp the desired features of the nominal beampattern, unless the array is exceptionally well calibrated. It should be noted that the expected beampattern is not a proper beampattern, but rather the statistical average of all possible realizations.

Although Gaussian error distributions may be a reasonable assumption in many situations, it can also give misleading results. When statistical assumptions, such as error independence, does not hold, sidelobe levels that should be statistically impossible may occur frequently. Moreover, a comprehensive statistical description may not be available or lead to an intractable formulation. Also, with an unbounded distribution, the beampattern will in principle be unbounded too. For these reasons, the statistical methods typically do not provide rigorous and finite upper and lower bounds on the beampattern.

Interval methods are generally suitable in various contexts that involve quantities that are bounded\citep{moore_introduction_2009, jaulin_applied_2001}. This is a weak restriction, as the quantities need not be precisely known or representable to be enclosed by an interval and give reliable results. In contrast to statistical methods, IA provides finite beampattern bounds given finite error bounds and weaker assumptions. 

The upper beampattern bound may be interpreted as a \emph{worst-case} beampattern performance in terms of the sidelobe level. Notably, the upper beampattern bound is also not a proper beampattern since it cannot be attained simultaneously in all directions. Because controlling the sidelobe level is a fundamental objective in array design, it is important to understand how tolerance errors in multiple variables can affect the beampattern; for instance, to mitigate the worst-case scenarios related to a high peak sidelobe level (PSLL). An earlier example of worst-case analysis is for phase quantization sidelobes\citep{holm_analysis_1992}. A comprehensive overview of calibration errors and analysis methods for phased array antennas can be found in the works by He et al.\citep{he_impact_2021}

The first application of IA in beampattern analysis was made in the antennae community by Anselmi et al.\citep{anselmi_tolerance_2013}, where they studied the effects of bounded amplitude errors. Subsequent works expanded the scope further. Poli et al.\citep{poli_dealing_2015} studied the effects of phase errors, whereas Zhang et al.\ studied joint amplitude and phase errors\citep{zhang_tolerance_2017}. Interval errors in the positions have also been investigated in various ways, such as in the case of bump-like features in reflector antennas\citep{rocca_interval-based_2014}. In beampattern analysis, the intervals reside in the complex plane. In the aforementioned works, the complex intervals are represented as rectangles in the complex plane (rIA, rectangular interval arithmetic).

Bounds resulting from mutual coupling errors have been analyzed for phased antenna arrays using the circular interval representation (cIA, circular interval arithmetic)\citep{anselmi_power_2016}. A similar analysis is based on the Cauchy-Schwarz inequality\citep{schmid_effects_2013}. However, in the mathematical models used therein, all error types are treated as special cases of mutual coupling errors, which blurs any clear separation between the different error types.

The rectangular and circular descriptions tend to overestimate the interval bounds by decoupling the inherent dependencies between the real and imaginary components. In order to produce tighter and more correct bounds, Tenuti et al.\citep{tenuti_minkowski_2017} proposed a polygonal representation (pIA, polygonal  interval arithmetic) by using Minkowski summation. To date, this is the most accurate method in the literature for this specific application. Other techniques, such as the Taylor-based interval method\citep{hu_new_2017}, also exist.

Recently IA has been introduced to sonar beamforming, starting as a cross-pollination from the antenna field. In the previous works on IA for phased-array antennas, a uniform linear array geometry was used. To make the theory more applicable to a wider range of sonars, Kirkeb{\o} and Austeng\citep{kirkebo_amplitude_2021} derived interval bounds for arrays of arbitrary shape and directive elements subject to amplitude errors by employing rIA. We have recently extended this framework and released a toolbox for beampattern interval analysis\citep{arnestad_sonar_2022}, which takes into account errors in amplitude, phase, position, and directivity by employing the tighter and more accurate pIA.

To the best of the authors' knowledge, there have been relatively few previous works published on IA in the context of acoustics. One example is the calculation of room acoustic reverberation times $T_{60}$ from bounded quantities such as volumes and sound absorption coefficients \citep{batko_uncertainty_2012}. Interval analysis has also been used to find all system configurations consistent with a set of measurements, as applied to underwater acoustic source localization\citep{brateau_acoustic_2022}. 

The current work builds upon Ref.~\citenum{arnestad_sonar_2022}, aiming to provide a more thorough analysis of worst-case situations using IA. To this end, the framework is extended to also include coupling, and describe it separately from the other forms of calibration errors. The key result of this study follows with ``backtracking'', which directly recovers the errors that result in a specific upper or lower bound beampattern magnitude. Backtracking provides insight into particularly unwanted error patterns that may result in exceptionally high PSLLs. The non-uniqueness of the bounds due to ambiguities in the error distribution (i.e., phase and position errors) is presented, along with a solution for resolving these ambiguities. Finally, an expression for the approximate beampattern bounds is derived. The expression provides insight into the limitations of array length and apodization for reducing the worst-case PSLL. It also sheds light on the similarities between the worst-case and expected beampatterns. The code used for this article is available online\cite{arnestad_beampattern_2022}.

The article is structured as follows: Sec.~\ref{sec:theory_background} covers the theoretical background, which primarily concerns beampatterns and real and complex interval arithmetic. Sec.~\ref{sec:formulation} presents the mathematical model to obtain bounded beampatterns with IA, while Sec.~\ref{subsec:direct_verification} introduces backtracking. In Sec.~\ref{sec:stats} an approximate bound is derived. All proposed methods are showcased as numerical experiments in Sec.~\ref{sec:results}. The results are discussed in Sec.~\ref{sec:discussion} and the article is concluded in Sec.~\ref{sec:conclusions}.


\section{\label{sec:theory_background} Theoretical background}

\subsection{\label{subsec:beampatterns} Beampatterns} 
Beamforming is the process of spatially filtering the wavefield using a sensor array. The array consists of $M$ elements at positions $\boldsymbol{r}_m$. The beamformer output is obtained by summing the appropriately delayed and weighted element inputs. The complex beampattern $B(\theta)$ describes the array's characteristics. In the far-field narrowband situation it is
\begin{equation}
\label{eq:nominal_beampattern}
    B(\theta) = \sum_{m=1}^{M} w_m \cdot d(\alpha_m) \cdot e^{j(\boldsymbol{k}(\theta) - \boldsymbol{k}_s)\cdot \boldsymbol{r}_m},
\end{equation}
for a wavefield arriving from the direction $\boldsymbol{k}(\theta)$ and steering direction $\boldsymbol{k}_s$. The array apodization is given by the element weights $w_m$ and allows for a trade-off between a narrower mainlobe and a decreased sidelobe level. In this work, the weights are always normalized such that $\sum_m w_m = 1$. The angular dependence is due to the relation
\begin{equation}
    \boldsymbol{k}(\theta) = [k_x, k_y]^\mathsf{T} = \frac{2 \pi}{\lambda} \cdot [\sin{\theta}, \cos{\theta}]^\mathsf{T},
\end{equation}
where $\lambda$ is the wavelength. Two-dimensional beamforming is considered, with arrays in the $x$-$y$ plane and broadside in $\boldsymbol{\hat{y}}$ the direction. The element position is a function of the coordinates $\boldsymbol{r}_m = [x_m, y_m]^\mathsf{T}$. 

\begin{figure}[t]
\includegraphics[width=0.8\reprintcolumnwidth]{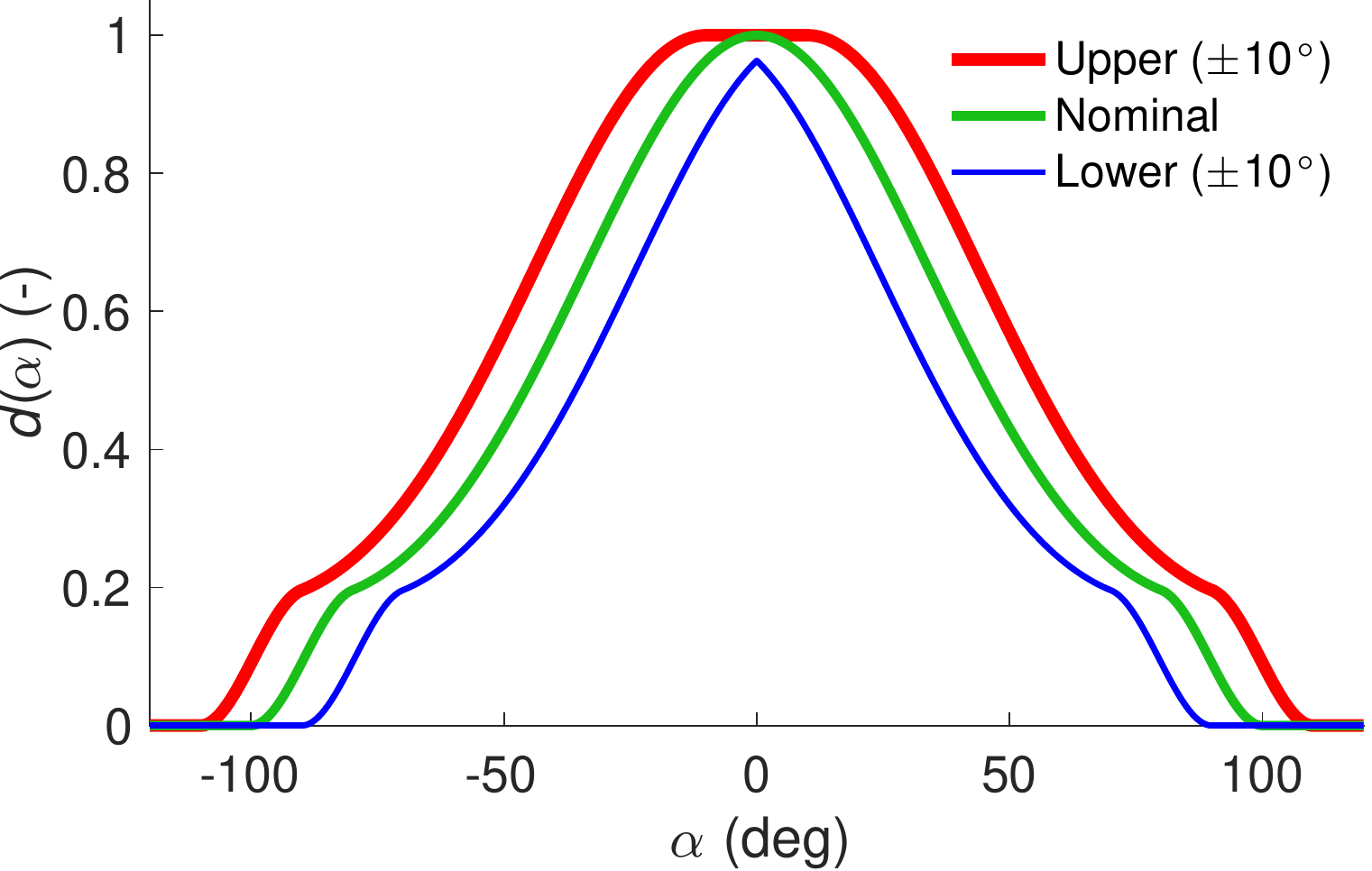}
\caption{{(Color online) The directivity function in Eq.~\eqref{eq:directivity} is evaluated within an orientation interval ($\alpha$-axis) of $\pm10^\circ$. $D = \lambda/2$, with tapered response between $80^\circ$ and $100^\circ$.}}\label{fig:directivity_function}
\end{figure}

The element-to-wavefield angle $\alpha_m = \theta - \psi_m$ determines the influence of the element directivity through the directivity function $d(\alpha_m)$. Here, $\psi_m$ is the angle orthogonal to the surface of the $m^{\text{th}}$ element. For circular elements of diameter $D$, the directivity takes on the form of the first-order Bessel aperture smoothing function \citep{johnson_array_1993}:
\begin{equation}
    \label{eq:directivity}
    d(\alpha) = 2 \cdot \jinc \left(\frac{2\pi \sin(\alpha)}{\lambda} D \right). 
\end{equation}
 The directivity function may be tapered to zero beyond 90$^\circ$ so that the element is not sensitive to the rear, see Fig.~\ref{fig:directivity_function}.

\begin{figure}[tb]
\includegraphics[width=1\reprintcolumnwidth]{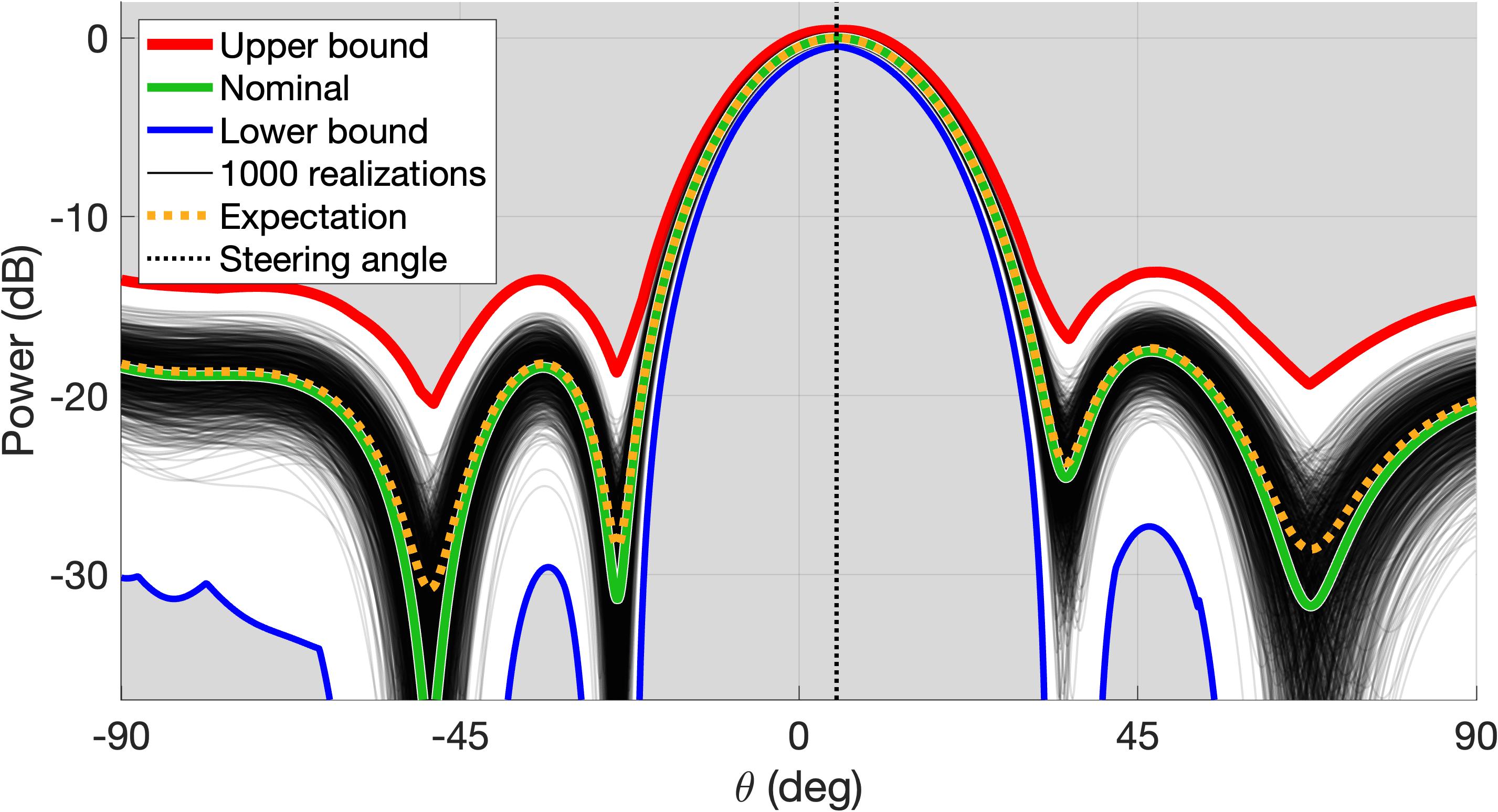}
\caption{\label{fig:beampatterns_plot}{(Color online) Beampatterns of example array A. The white region signifies the area between the bounds.}}
\end{figure}

It is customary to mainly consider the output power, defined as $P(\theta) = |B(\theta)|^2$. In Fig.~\ref{fig:beampatterns_plot}, the nominal error-free beampattern for an $M=5$ element curved\citep{kirkebo_amplitude_2021} sonar is illustrated, as obtained using Eq.~\eqref{eq:nominal_beampattern}. The array parameters are specified in Table~\ref{tbl:arrayA}. This example is meant to be illustrative and is referred to as array example A. The speed of sound throughout the text is assumed to be $1500 \, \mathrm{m/s}$, and the wave frequency is set to $20 \, \mathrm{kHz}$. 

\begin{table}[ht]
\vspace{-0.2cm}
\caption{Example array A.}
\vspace{-0.3cm}
\label{tbl:arrayA}
\centering
\begin{ruledtabular}
\begin{tabular}{l r}
Number of elements, $M$ & 5   \\ 
Element pitch & $\lambda/2$   \\
Element diameter, $D$   & (omnidirectional) $\approx 0 \lambda$ \\
Radius of array curvature   & $8/3\lambda$     \\
Apodization, $\mathbf{w}$  & $[14, 23, 27, 23, 14] \%$  \\
Steering angle, $\theta_s$ & $5^\circ$ \\
Maximum amplitude error, $\delta g$ & $\pm 5\%$ \\
Maximum phase error, $\delta \mathbf{\Phi}$ & $\pm [6, 4.5, 4, 4.5, 6]^\circ$
\end{tabular}
\end{ruledtabular}
\vspace{-0.2cm}
\end{table}

The beampattern bounds, using the error intervals given in Table~\ref{tbl:arrayA}, are introduced briefly in this section. The bounds are seen in Fig.~\ref{fig:beampatterns_plot}, and the method of calculation is outlined in Sec.~\ref{sec:formulation}. Between the two bounds, 1000 random realizations are plotted, illustrating the inclusive property of IA. The errors are drawn uniformly and independently from the intervals. Non-uniform phase bounds are chosen to highlight that edge elements may have different neighboring conditions, and that the IA framework can handle element-dependent error sizes. For instance, this could be relevant for thermal expansion where element positional deviation is proportional to the distance from the attachment point. 

Taking a statistical approach, it is found that the expected power for Gaussian amplitude and small phase errors\citep{van_trees_optimum_2004} is approximately
\begin{equation}
\label{eq:expected_P}
     \mathbb{E}\left\{ P(\theta) \right \} \approx   |B(\theta)_{\mathrm{nom.}}|^2 \cdot e^{-\sigma_{\phi}^2 }  
      + T_\text{se} \cdot \left( \sigma_g^2 + \sigma_{\phi}^2 \right),
\end{equation}
where $\sigma_{g}^2$ and $\sigma_{\phi}^2$ are the variances for element amplitude and phase. $T_\text{se}$ is the sensitivity function, defined as 
\begin{equation}
    T_\text{se} = \sum_{m=1}^M |w_m| ^2.
\end{equation}
In deriving this expression, we make the assumption that no variations depend on $\theta$. The second term in Eq.~\eqref{eq:expected_P} raises the power uniformly, affecting the ability to specify beampattern nulls in particular. The expected beampattern is also shown in Fig.~\ref{fig:beampatterns_plot}. The variances for the non-uniform phase errors are calculated by averaging the variance across the elements, assuming uniform distributions. In Sec.~\ref{sec:stats}, we derive an expression for the worst-case beampattern, showing some resemblance in its formulation with the expected beampattern.

\subsection{\label{subsec:2:1} Interval arithmetic of real numbers}

Interval variables are indicated with the superscript $I$ and represent a connected set of numbers: 
\begin{equation}
x^I = [\underline{x}, \overline{x}] = \{x \in \mathbb{R} : \underline{x} \leq x \leq \overline{x} \},    
\end{equation}
where $\underline{x}$ and $\overline{x}$ are the lower and upper interval bounds, respectively. As with ordinary variables, operations can be performed on intervals. For the addition of two intervals
\begin{equation}
    x^I + y^I = [\underline{x} + \underline{y}, \overline{x} + \overline{y}],
\end{equation}
and for multiplication
\begin{equation}
\begin{split}
        x^I \cdot y^I = [ & \text{min}\left\{\underline{x}  \underline{y}, \underline{x}  \overline{y}, \overline{x}  \underline{y}, \overline{x}  \overline{y}\right\}, 
                   \text{max}\left\{\underline{x}  \underline{y}, \underline{x}  \overline{y}, \overline{x}  \underline{y}, \overline{x} \overline{y} \right\} ].
\end{split}
\end{equation}

An important feature of interval arithmetic is that subtraction and division are not additive or multiplicative inverses, except in the special case of degenerate intervals. In other words, $x^I - x^I \neq [0,0]$, unless the upper and lower bounds are equal.

The interval output of functions can also be defined. For example, consider the quadratic function $f(x) = x^2$, which has a minimum at $x=0$. This highlights the importance of checking if an interval contains any extrema of the function. For the interval $x^I$, the function $f(x^I) = \{x^2 : x \in x^I \} $ can be expressed as
\begin{equation}
\label{eq:function_interval}
    f(x^I) =
    \begin{cases}
      [\overline{x}^2, \underline{x}^2] & \text{if $\overline{x} \leq 0$},\\
      [0, \text{max} \left\{ \underline{x}^2, \overline{x}^2 \right\} ] & \text{if $0 \in x^I$}, \\
      [\underline{x}^2, \overline{x}^2] & \text{if $\underline{x} \geq 0$}.
    \end{cases} 
\end{equation}

The final issue to address in interval arithmetic is \emph{the dependence problem}. This occurs when a variable is represented more than once in an expression. For example, if the interval $x^I = [\underline{x}, \overline{x}] = [-1,1]$ is naively multiplied with itself, the result would be $x^I \cdot x^I = [-1, 1]$ instead of $(x^I)^2 = [0,1]$, because the $x^I$ in the first factor is treated independently from the second factor. This is due to the lack of distributivity, but it can also result from the lack of additive and multiplicative inverses in interval arithmetic \citep{moore_introduction_2009}.

\subsection{\label{subsec:array_errors} Array errors and complex interval arithmetic}

\begin{figure}[tb]
\includegraphics[width=0.6\reprintcolumnwidth]{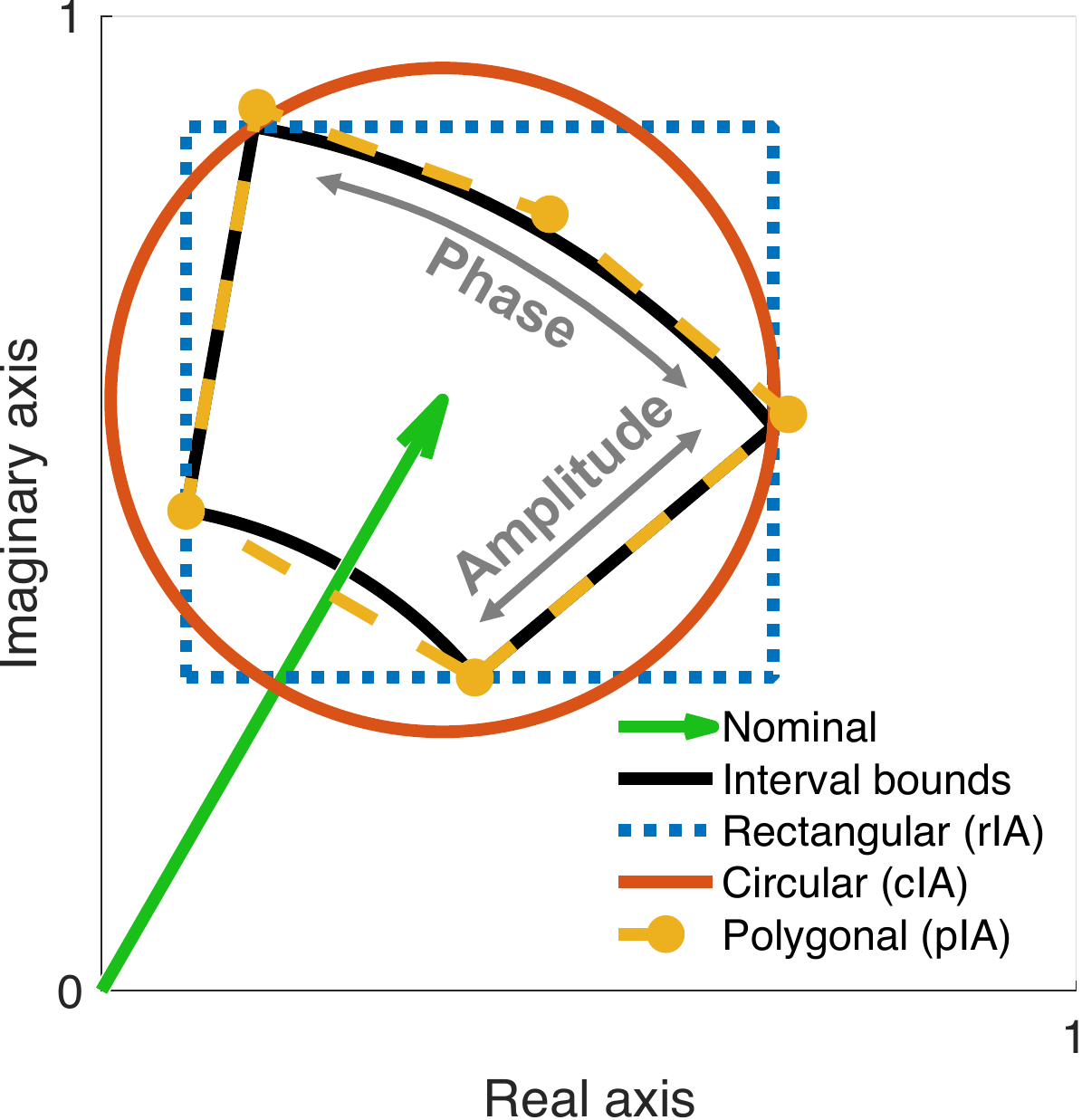}
\caption{\label{fig:interval_representation}{(Color online) The nominal element response is indicated by the arrow tip. Bounded errors in amplitude and phase give interval bounds that are shaped like annular sectors. Alternative representations, like the polygonal wrapping, are used to enable arithmetic operations.}}
\end{figure}

If the array elements are subject to bounded errors in phase and amplitude, the phasor values in Eq.~\eqref{eq:nominal_beampattern} are bounded within a two-dimensional shape in the complex plane known as an annular sector, as seen in Fig.~\ref{fig:interval_representation}. This annular sector is considered to be a two-dimensional complex interval.
In the beamforming process, the complex intervals are summed, resulting in a complex interval $B^I$ as the output. Mathematically, for two intervals $A_1^I$ and $A_2^I$, the Minkowski sum is defined as the set of all possible sum combinations
\begin{equation}
    A_{\text{sum}}^I = A_1^I + A_2^I = \{ A_1 + A_2 : A_1 \in A_1^I, A_2 \in A_2^I\}.
\end{equation}
How this sum is performed in practice depends on the chosen interval representation, of which some examples are shown in Fig.~\ref{fig:interval_representation}. Rectangular and circular intervals enclose the annular sectors and were the first used for beampattern analysis\citep{anselmi_tolerance_2013, anselmi_power_2016}. While these representations are convenient for summation, they are evidently not tight and can introduce ``pessimism'' to the bounds, also known as \emph{the wrapping problem} in IA.

A later development in this field involved integrating IA with Minkowski summation by wrapping the annular sectors with convex polygons\citep{tenuti_minkowski_2017}, as seen in Fig.~\ref{fig:interval_representation}. The inner, concave part of the annular sector is included by forming the convex hull. Although this might seem problematic, the Shapley-Folkman lemma shows that the Minkowski summation is a convexifying operation. In other words, the sums of many concave sets are approximately convex\citep{schneider_convex_1993}. This deviation can be quantified with the Hausdorff distance, but this topic is not a major concern because the bounds are inclusive in any case.

Considering only the convex boundary is sufficient because Minkowski summation and forming the convex hull are commuting operations \citep{schneider_convex_1993}. The sampling resolution of the curve is determined by the error tolerance in the representation, as the polygon must enclose the arcs of the annular sectors. A summation algorithm can be implemented that runs in linear time with respect to the number of vertices on the two boundaries\citep{de_berg_computational_2008}, based on comparing only vertex pairs that are extreme in the same direction.

Throughout this article, other complex interval operations are needed. For example, the absolute value of a complex interval is needed to plot the power bounds $\underline{P}(\theta)$ and $\overline{P}(\theta)$. This real-valued interval gives the distance from the origin to the closest and furthest point on the boundary of $B^I$. If the origin is contained within the interval, then the minimum distance is zero. Additionally, multiplication of complex intervals is only needed for a special case, which is discussed in Sec.~\ref{subsec:coupling_interval}.

\section{\label{sec:formulation} Beampattern bound formulation}

\subsection{\label{subsec:error_model} Element error model}

\begin{figure}[tb]
\includegraphics[width=0.75\reprintcolumnwidth]{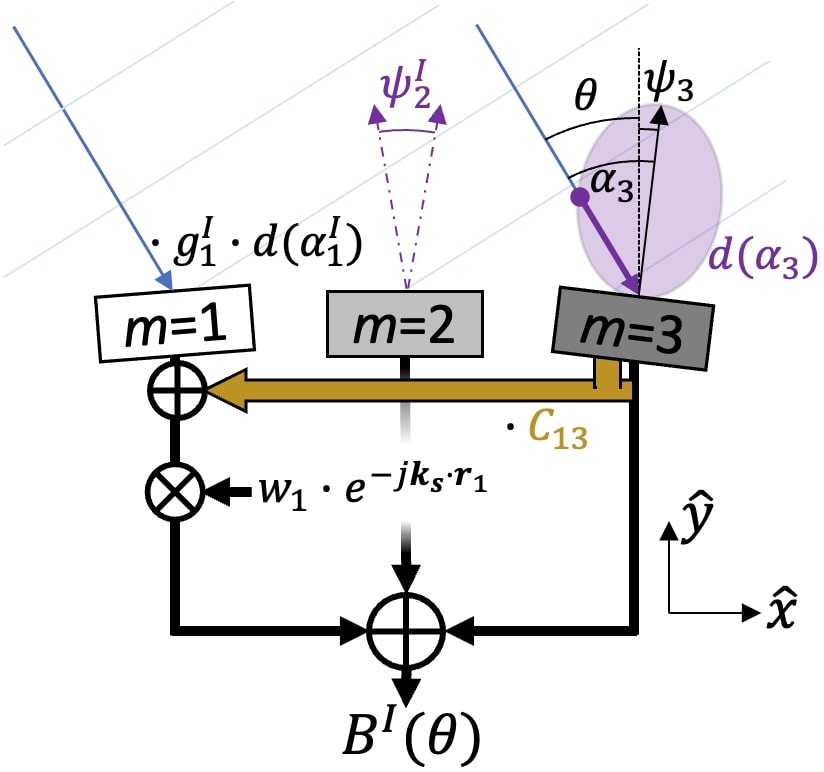}
\caption{(Color online) Error model exemplified with a 3-element array. Each element is subject to bounded errors and coupling prior to electronic steering and apodization. Phase and position errors apply to each element, but are not shown.}
\label{fig:error_model}
\end{figure}

The model for element errors and the connection to the beampattern bounds are derived with reference to Eq.~\eqref{eq:amp_phase_B_int} and Fig.~\ref{fig:error_model}, where a plane wave is incident to the array. Note that no assumptions are made about the array geometry, and that coupling is treated separately in Sec.~\ref{subsec:coupling_interval}. 

Errors are allowed in the element amplitude $g_m$, the element directivity $d(\alpha_m)$, the element positions $\boldsymbol{r}_m$, and the phase $\Phi_m$, so that they lie within certain intervals, such as $g_m^I = \left[ \underline{g}_m, \, \overline{g}_m \right]$.
%
%
Typically, the intervals are specified to be symmetric around some nominal value, referred to as the interval mid-point. Let $\delta$ denote the maximum error (or half of the interval width), so for example, $g_m^I = \left[ 1- \delta g_m, \, 1+ \delta g_m \right]$. Note that $\boldsymbol{r}^I_m$ is taken to be a vector with independent interval components, resulting in rectangular areas around the nominal element positions. These intervals can be directly inserted into Eq.~\eqref{eq:nominal_beampattern} to give
\begin{equation}
\label{eq:amp_phase_B_int}
\begin{split}
    B^I(\theta) &= \sum_{m=1}^M  \underbrace{w_m \cdot g_m^I \cdot d_m(\alpha_m^I) }_{\text{Amplitude interval: } a_m^I }\cdot  \underbrace{e^{j\left( \boldsymbol{k}(\theta) \cdot \boldsymbol{r}^I_m + \Phi_m^I - \boldsymbol{k}_s\cdot \boldsymbol{r}_m \right)}}_{\text{Phase interval: } \varphi_m^I}.
\end{split}
\end{equation}
Note that the steering is applied electronically under the assumption that element positions are known to be $\boldsymbol{r}_m$, as opposed to the actual element positions that may be within the intervals $\boldsymbol{r}^I_m$. 

This formulation allows for the use of real-valued IA for the amplitude intervals $a_m^I$ and phase intervals $\varphi_m^I$ separately, making it manageable to work with multiple bounded categories of error. Unlike earlier descriptions that have focused on deriving analytical bounds\citep{kirkebo_amplitude_2021}, the expression stays close to the original formula for calculating beampatterns.

The directivity function $d(\alpha)$ in Eq.~(\ref{eq:directivity}) is evaluated with interval inputs. By assuming a tapered aperture smoothing function with one global maximum, the same logic as demonstrated in Eq.~\ref{eq:function_interval} can be used in the evaluation. The interval function used in this work is shown in Fig.~\ref{fig:directivity_function}.



\subsection{Element-to-element coupling}\label{subsec:coupling_interval} 

Mutual coupling, also known as cross-talk, is now introduced into the formulation of Eq.~(\ref{eq:amp_phase_B_int}). This refers to the transfer of signal from one element to the output line of another element, either through electrical or mechanical mechanisms. This is shown in Fig.~\ref{fig:error_model}, where an arrow from element $c=3$ to $m=1$, signifies that the \emph{coupling} element $c=3$ is multiplied with a coefficient $C_{mc}$ and injected into the channel of element $m=1$. These mechanisms are often expressed using matrix formulations.

In this study, coupling coefficients are treated as unknown but bounded complex intervals $C_{mc}^I$. Estimating reliable bounds on these coefficients is challenging, as the exact coupling model is generally not readily available\cite{anselmi_power_2016}. Therefore, it is assumed that the coupling phase is completely unknown, and this assumption cannot easily be reversed in the following derivation. The maximum coupling magnitude, on the other hand, is taken to decrease exponentially away from the excited element (this is not a necessity in the following derivation):
\begin{subequations}
    \begin{align}
        \left|C^I_{mc}\right| &= \left[0, \gamma^{|m-c|} \right], \\ 
        \angle C^I_{mc} & = [0, 2\pi \cdot (1-\delta_{mc})].
    \end{align}
\end{subequations}
Here $\gamma$ is the magnitude of the element-to-element coupling. The phase is unknown, except in the case of self-coupling $C_{mm} = 1$ when $m=c$. This description essentially treats coupling as circular intervals, which previous coupling models have also done \citep{anselmi_power_2016, schmid_effects_2013}. However, our model maintains a meaningful separation between element errors and coupling errors, which allows for using the tighter polygonal representation for the main element intervals.

To incorporate the effects of coupling, it is intuitive to write out the model depicted in Fig.~\ref{fig:error_model}, expanding upon Eq.~\eqref{eq:amp_phase_B_int}. The beamformer output is as always a sum over the index $m$, but the contributions from each element $c$ that couples into channel $m$ must also be included:
\begin{widetext}
\begin{equation}
\label{eq:AP_intervalbeampattern_coupled_original}
    B^I(\theta) = \sum_{m=1}^M w_m \cdot e^{-j \boldsymbol{k}_s\cdot \boldsymbol{r}_m} \cdot \left( \sum_{c=1}^M  |{C}|^I_{mc} \cdot  g_c^I \cdot d_c(\alpha_c^I) \cdot e^{j(\angle {C}^I_{mc} +\boldsymbol{k}(\theta) \cdot \boldsymbol{{r}}^I_c + {\Phi}_c^I)}  \right) .
\end{equation}
In the absence of coupling, the expression reduces to Eq.~\eqref{eq:amp_phase_B_int}. However, due to the outer sum, each interval (such as $g_c^I$) is evaluated $M$ times independently. To reduce this dependence problem, the two sums can be swapped:
%
\begin{equation}
\label{eq:interval_BP}
    B^I(\theta) = \sum_{c=1}^M g_c^I \cdot d_c(\alpha_c^I) \cdot e^{j(\boldsymbol{k}(\theta) \cdot \boldsymbol{r}^I_c + \Phi_c^I)} \left( \sum_{m=1}^M |C|^I_{mc} \cdot w_m \cdot e^{j(\angle C^I_{mc} - \boldsymbol{k}_s\cdot \boldsymbol{r}_m)}\right) .
\end{equation}
\end{widetext}
This formulation provides tighter bounds, and the terms are also conveniently grouped by initially summing over circular intervals, allowing for the use of the following instrumental notation
\begin{subequations}
\label{eq:E_and_A}
\begin{align}
    E_{c}^I &= g_c^I \cdot d_c(\alpha_c^I) \cdot e^{j(\boldsymbol{k}(\theta) \cdot \boldsymbol{r}^I_c + \Phi_c^I)} = a_c^I e^{j \varphi_c^I}, \label{eq:E_and_A_E} \\ 
    A^I_{c} &= \sum_{m=1}^M |C|^I_{mc} \cdot w_m \cdot e^{j(\angle C^I_{mc} - \boldsymbol{k}_s\cdot \boldsymbol{r}_m)}, \label{eq:E_and_A_A}
\end{align}
\end{subequations}
so that the beampattern intervals with coupling can be written as 
\begin{equation}
\label{eq:AP_intervalbeampattern_coupled_simplified}
    B^I(\theta) = \sum_{c=1}^M E_c^I \cdot A^I_c.
\end{equation}
Here $E_{c}^I = a_c^I \cdot e^{j \varphi_c^I}$ represents a complex annular sector interval that \emph{only} describes an element. On the other hand, $A^I_{c}$ is a complex circular interval that determines the element's interaction with the array structure, including effects such as coupling, apodization, and steering. The circle is centered at $w_c \cdot e^{- j\boldsymbol{k}_s\cdot \boldsymbol{r}_c}$ and has a radius of $ R_{Ac} = \sum_{\substack{m=1 \\ m \neq c}}^M \gamma^{|m-c|} \cdot w_m$.

\begin{figure}[tb]
\includegraphics[width=0.8\reprintcolumnwidth]{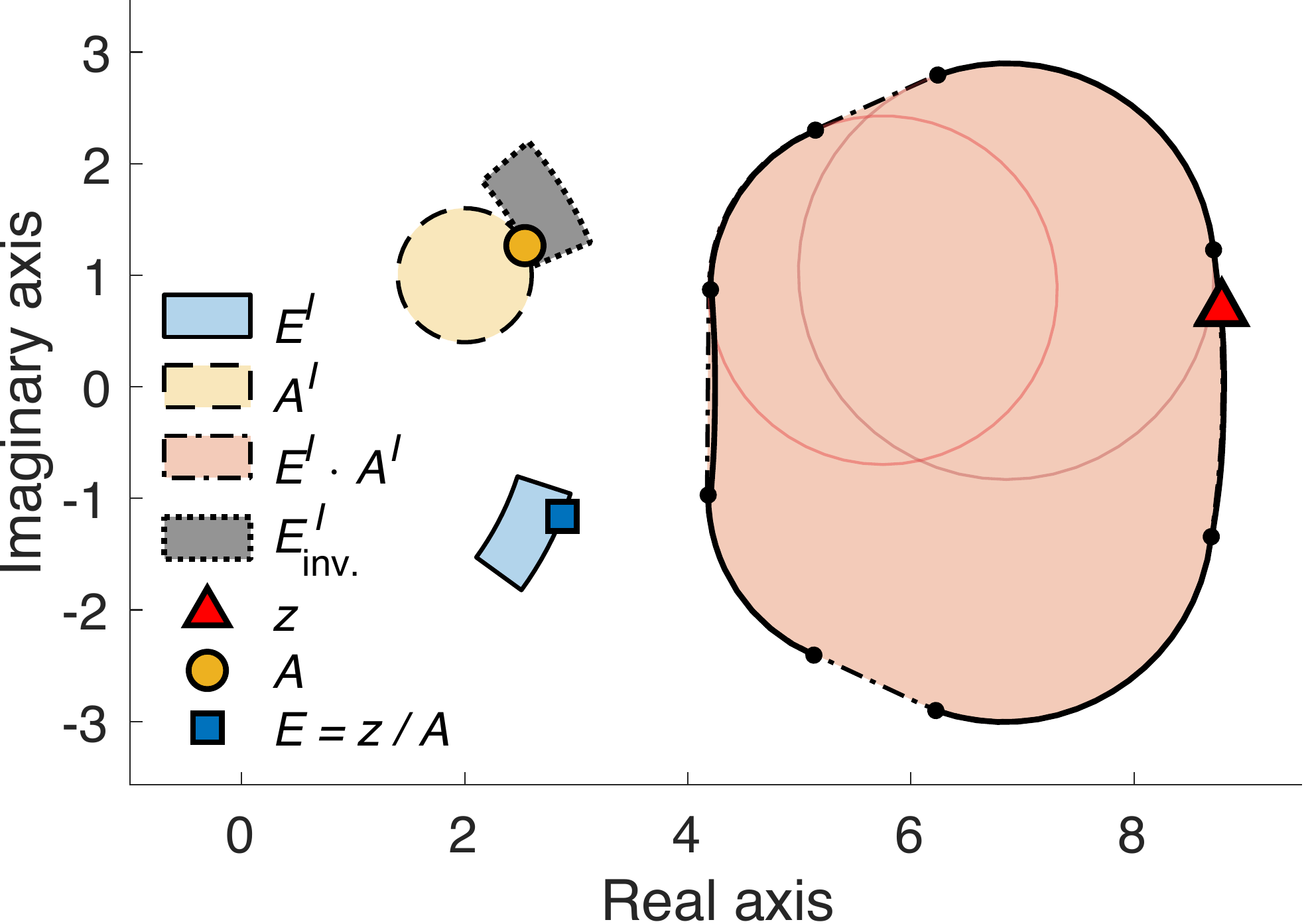}
\caption{\label{fig:coupling_backtrack}{ (Color online) The product and backtracking of various intervals, scaled for illustrative purposes and subscript $c$ omitted. $E_c^I \cdot A_c^I$ is a rounded annular sector. Points $z$ on this boundary can be uniquely backtracked into the factor intervals, so that $E_c \cdot A_c = z_c$. }}
\end{figure}

The product of $E_c^I \cdot A^I_c$ results in ``rounded'' annular sectors, as shown in Fig.~\ref{fig:coupling_backtrack}. These shapes are obtained through the complex Minkowski product\cite{farouki_minkowski_2001}, which can be understood as the union of many scaled and rotated circles. Upon closer examination, one can show that the product boundary consists of six circular arcs connected by linear segments; one for each of the four corners, and two for the inner and outer arcs of the annular sector. As discussed in Sec.~\ref{subsec:array_errors}, the convex boundary is used for the polygonal representation. Since the inner arc is concave, the convex rounded annular sector can be described using five arcs. Thus, the sum with coupling is performed over polygons that represent rounded annular sectors.


\section{\label{subsec:direct_verification}  Backtracking: direct bound verification}

\begin{figure*}[tb]
\figline{
\fig{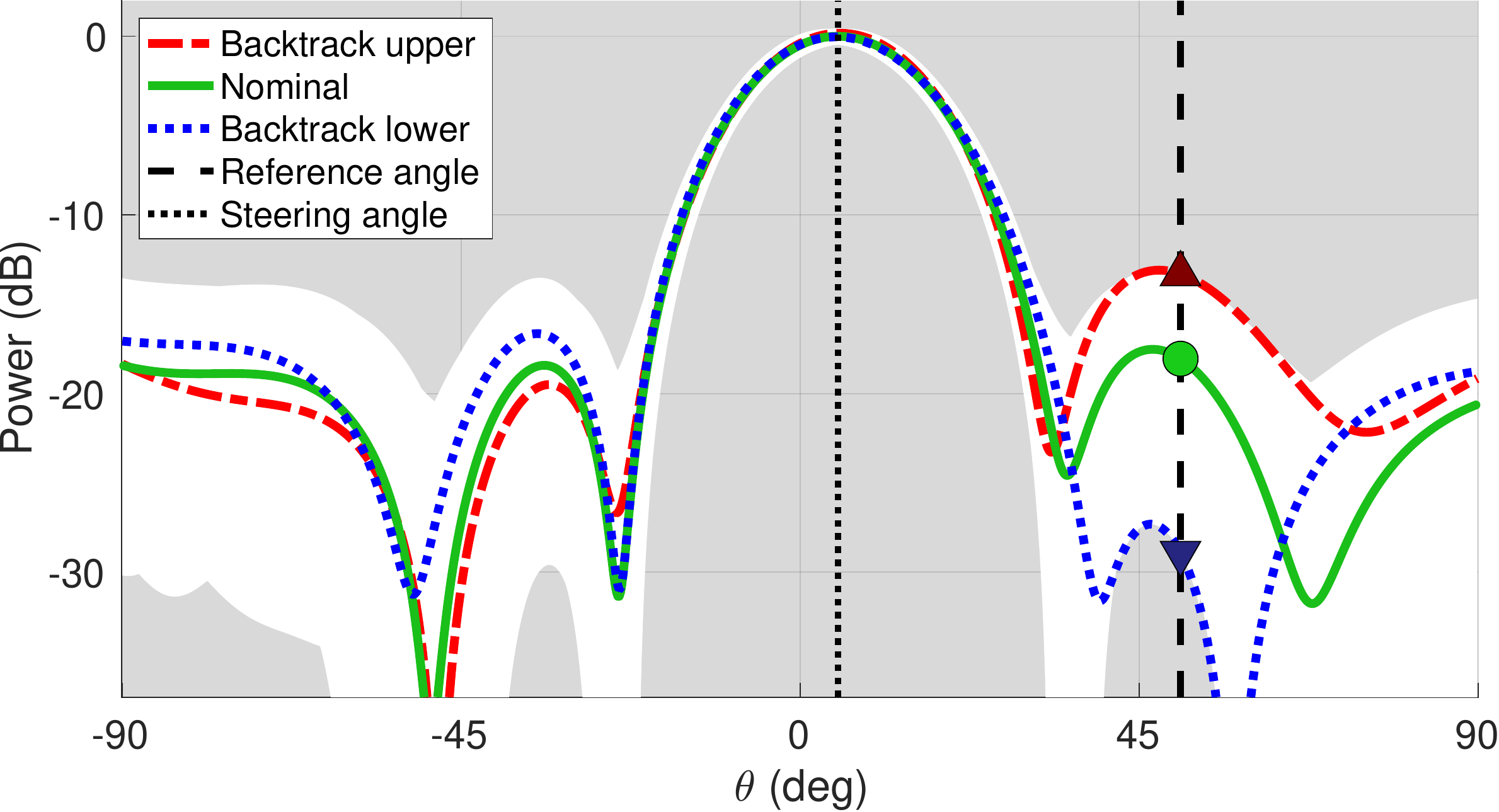}{.6\textwidth}{(a)}\label{fig:recovered_beampattern}
\fig{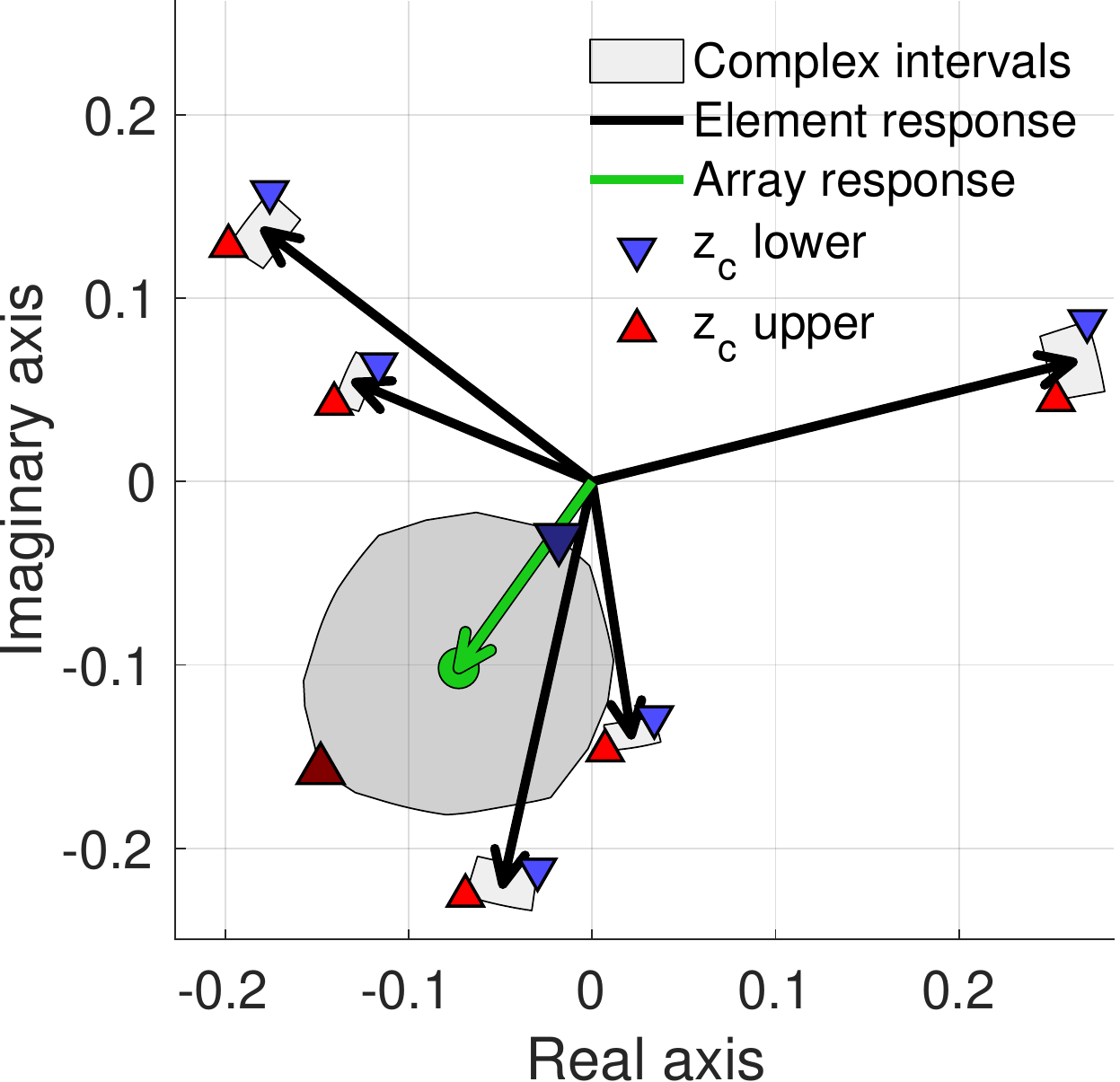}{.325\textwidth}{(b)}\label{fig:backtrack_shapes_and_potato}
}
\caption[]{(Color online) \label{fig:array_sum} Panel (a) shows the nominal power pattern, with the backtracked upper and lower bounds for example array A. Panel (b) shows how the array response is a sum of the complex-valued element responses. The bounds (triangles) are the most extreme values possible in the array response.}
\end{figure*}

The calculated beampattern bounds have in most previous works been verified using Monte Carlo simulations \citep{tenuti_minkowski_2017, kirkebo_amplitude_2021}. However, due to the statistical nature of such simulations, there is no guarantee of achieving the exact bounds. Alternatively, one could employ optimization methods to search for these bounds in a high-dimensional space.

In this section, we develop a novel technique for directly verifying the bounds. It works by recovering the errors corresponding to the beampattern that reaches the bound. This is illustrated in Fig.~\ref{fig:recovered_beampattern}, and refer to this technique as ``backtracking'' as the intention is to backtrack the contributing points in the complex intervals of the elements from the summed interval $B^I$.

\subsection{\label{subsec:backtrack_simple}Simple phase and amplitude intervals}

We first consider a situation with simple phase and amplitude intervals, and no directivity effects or positional errors (both involve dependence on $\theta$), or coupling. In that situation
\begin{subequations}
\label{eq:E_and_A_2}
\begin{align}
    E_{c}^I &= g^I_c \cdot e^{j(\boldsymbol{k}(\theta) \cdot \boldsymbol{r}_c + \Phi_c^I)}, \\ 
    A_{c} &= w_c \cdot e^{ -j \boldsymbol{k}_s\cdot \boldsymbol{r}_c}.
\end{align}
\end{subequations}
Next, we choose a reference angle $\theta_{\text{ref.}}$, for example $50^\circ$ as shown in Fig.~\ref{fig:recovered_beampattern}, to backtrack the specific errors $g_c \in g^I_c$ and $\Phi_c \in \Phi_c^I$ associated with either $\overline{P}(\theta_{\text{ref.}})$ or $\underline{P}(\theta_{\text{ref.}})$. For the sake of this argument, we choose $\overline{P}$, which is the sum of $M$ complex numbers $z_c$ (with $c = 1, \ldots, M$). These numbers $z_c$ are found on the boundaries of the respective intervals $z_c \in \partial (E_{c}^I \cdot A_c)$, and
\begin{equation}
\label{eq:chosen_P_upper}
    \overline{P}(\theta_{\text{ref.}}) = \left| \sum_{c=1}^M z_c \right|^2.
\end{equation}
This can be interpreted as the squared maximum distance from the beampattern response to the origin, as illustrated in Fig.~\ref{fig:backtrack_shapes_and_potato}. In order to perform the backtracking, the terms that go into the sum must be recovered, that is, solving
\begin{equation}
\label{eq:chose_p}
    \{z_1, \ldots, z_M \}  = \argmax_{z_c \in E_{c}^I \cdot A_c} \bigg|\sum_{c=1}^M z_c \bigg|^2.
\end{equation}

A method for efficiently recovering $\{z_1, \ldots, z_M \}$ can be implemented using the same principle as the linear time Minkowski sum algorithm. A detailed description of the algorithm can be found in Ch.~13 of Ref.~\citenum{de_berg_computational_2008}, although it is not a prerequisite for the following discussion. Following Fig.~\ref{fig:Gauss_map_matching}, the first step is to find the vertex in the summed polygon that is farthest the origin. The extreme direction will lie between the outer vertex normals. The key concept is that only pairs of vertices that are extreme in the same direction contribute to the Minkowski sum boundary, and therefore this must also apply to the backtracked vertices. For each polygon in the sum, one needs to look for the corresponding vertex $z_c$ whose outer normals contain this extreme direction. The normals can be found from the edges connecting each vertex to its neighbor. The same method and argument hold for the least extreme vertex by reversing the extreme vector direction. Since backtracking involves a linear search for the vertex satisfying the outer normal condition, the complexity of backtracking all polygons is $\mathcal{O}(M\cdot N_\text{vert.})$, where $N_\text{vert.}$ is the number of vertices used to sample each polygon.

\begin{figure}[tb]
\includegraphics[width=0.7\reprintcolumnwidth]{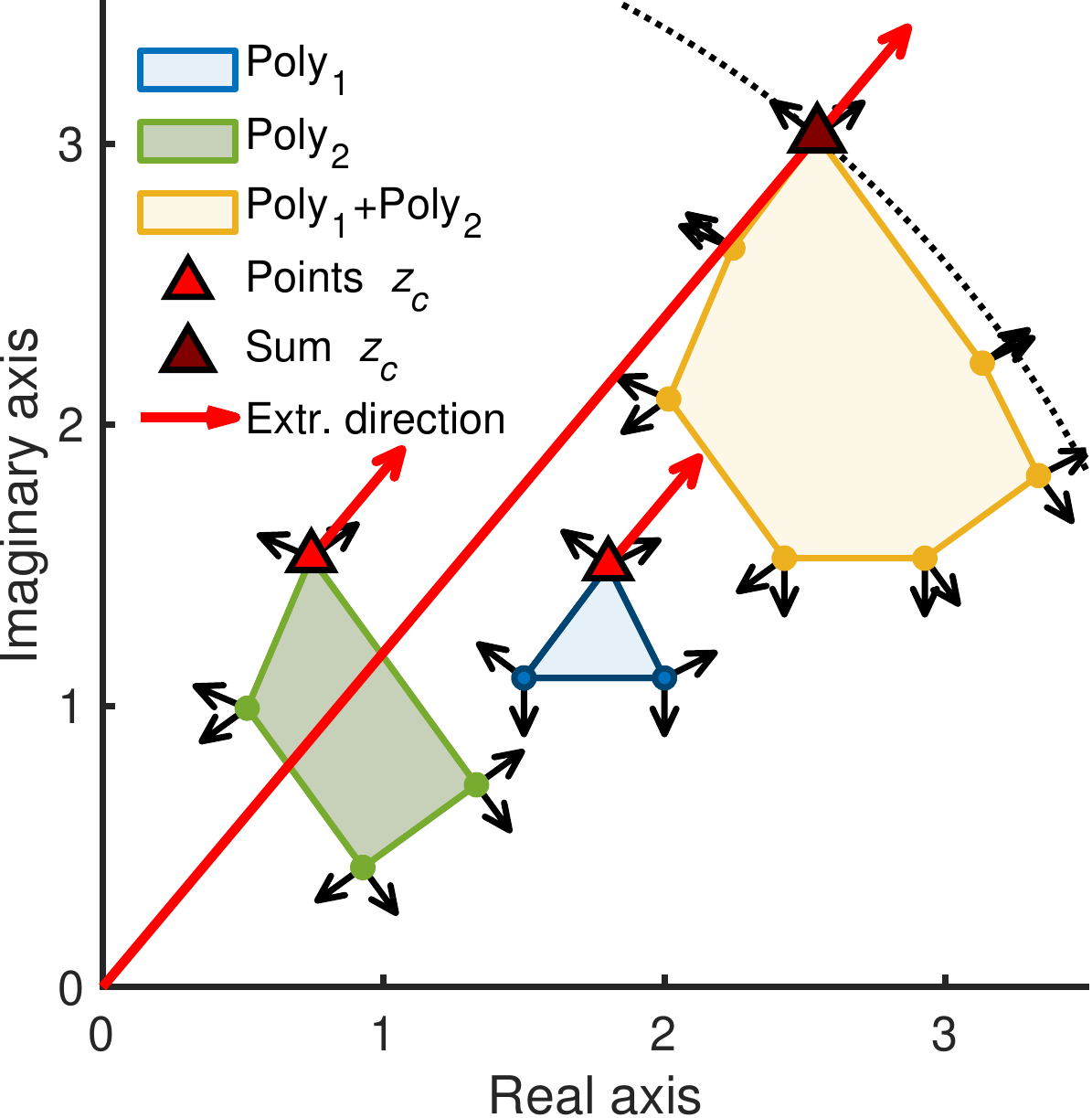}
\caption{\label{fig:Gauss_map_matching}{(Color online) The points $z_c$ that contribute to the most extreme vertex on the polygon sum can be found directly by matching the extreme direction with the outer vertex normals.}}
\end{figure}

\begin{figure}[tb]
\includegraphics[width=1\reprintcolumnwidth]{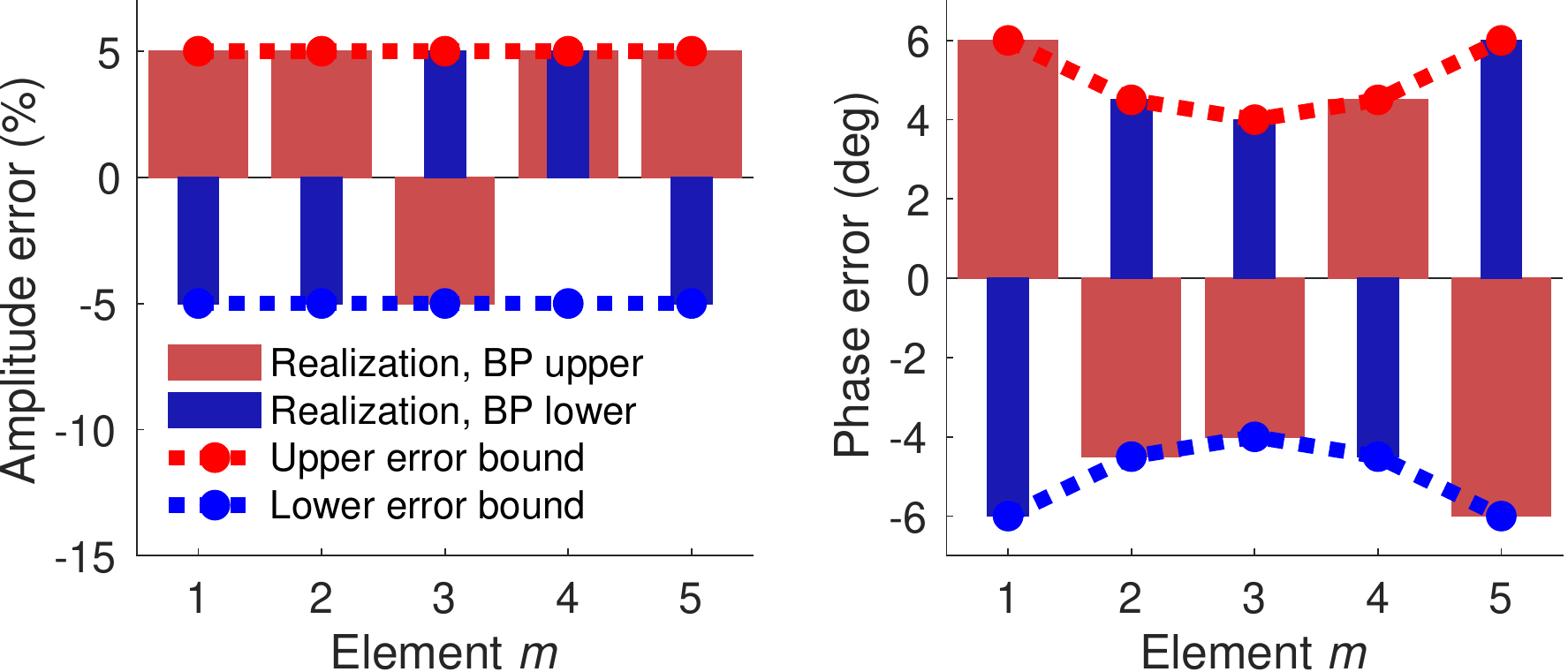}
\caption{\label{fig:recovered_errors}{(Color online) Amplitude (left panel) and phase (right panel) of the backtracked errors $\varepsilon$ for each element, corresponding to the upper bound in Fig.~\ref{fig:recovered_beampattern}. It is worth noting that the indices $c$ and $m$ can be used interchangeably here.}}
\end{figure}

With the points $z_c$ known, the element errors and the corresponding beampattern can be calculated. The element errors $\varepsilon_c = g_c \cdot e^{j \Phi_c}$ can be unambiguously obtained by undoing the phases from both steering and the wavefield, together with the apodization
\begin{equation}
\label{eq:elem_weights_no_k_or_C}
    \varepsilon_c = z_c / (A_c \cdot e^{- j\boldsymbol{k}(\theta_{\text{ref.}}) \cdot \boldsymbol{r}_c}).
\end{equation}
In Fig.~\ref{fig:recovered_errors}, the phase and amplitude realizations for the upper and lower beampattern bounds in Fig.~\ref{fig:recovered_beampattern} are shown, along with their respective error bounds. It should be noted that while the amplitude errors will always be extreme, in the sense that either the upper or lower error bounds are reached, this is not necessary for the phase. The plot also serves to verify how well the bounds imposed on the errors are respected in the construction of the bounded beampattern.


The beampattern is obtained by considering a wavefield impinging from another direction. Since the specific errors are known, one can simply apply them as weights when computing the beampattern
\begin{equation}
\label{eq:reconstructed_BP_no_k_or_C}
    B(\theta | \theta_{\text{ref.}} ) = \sum_{c=1}^M w_c \cdot \varepsilon_c \cdot e^{j (\boldsymbol{k}(\theta) - \boldsymbol{k}_s )\cdot \boldsymbol{r}_c}.
\end{equation}
In Fig.~\ref{fig:recovered_beampattern}, the unique beampatterns that reach the upper and lower bounds at the reference angle are depicted. It should be noted that certain features, such as the maximal mainlobe width, cannot be backtracked, since the bounds cannot be realized simultaneously in all directions.

\subsection{Non-unique bounds}

Backtracking becomes ambiguous under two circumstances, including cases when coupling is considered. Firstly, in rare instances, there may be two or more equally extreme vertices to choose between. Secondly, and more relevant, when positional errors and phase response errors contribute to the combined phase error, and a non-extreme combined phase error is required to reach the beampattern bounds. This occurs when the extreme direction falls within the opening angle of an annular sector. In the following it is shown that backtracking can still be performed with phase and position errors jointly, as long as ambiguities are resolved when they occur, by deciding how the error is distributed over the different variables. To simplify the argument without loss of generality, assume that the position error is only in dimension $x$. 

The interval of \emph{possible} phase values $z_c$ can assume is $\angle z_c^I = k_x x^I + \Phi^I$. If the phase angle of $z_c$ is not at the interval bounds, one must decide how the phase error is distributed among $x$ and $\Phi$. In this case, the choice is made to decide the error in $\Phi$ first. Denote the selected value as $\Phi^*$. The maximum value that can possibly be selected is indicated as $\overline{\Phi^*}$, and cannot be greater than $\overline{\Phi}$ under any circumstances. At the same time, it is limited by $\angle z_c$ and the lower bound of $k_x x^I$. Thus, any values of $\Phi^*$ falling within the following bounds are valid
\begin{subequations}
\label{eq:ambiguity}
\begin{align}
     \overline{\Phi^*} &= \min \left[ \overline{\Phi}, \angle z_c - \underline{k_x x} \right],\\ 
     \underline{\Phi^*} &= \max \left[ \underline{\Phi}, \angle z_c - \overline{k_x x} \right].
\end{align}
\end{subequations}
After choosing the value of $\Phi^*$, the required value of $x^*$ will be given as 
\begin{equation}
    x^* = \frac{\angle z_c - \Phi^*}{k_x}.
\label{eq:second_ambiguity_resolv}
\end{equation}
This way of resolving the ambiguities can be repeated when more errors contribute to the phase. As a consequence of these ambiguities, there may be multiple ways to realize the upper beampattern bound.



\subsection{With coupling}
If coupling (Sec.~\ref{subsec:coupling_interval}) is included, the backtracking is further complicated because the complex values can be taken from anywhere on the boundary of the product of two intervals, $z_c \in \partial (E_c^I \cdot A_c^I)$. The algorithm described in Sec.~\ref{subsec:backtrack_simple} can be used to find these values, but with the additional requirement of determining which value in $E_c^I$ and $A_c^I$ could result in that particular $z_c$, and whether that value is unique or not.

Fig.~\ref{fig:coupling_backtrack} displays the product of $E_c^I \cdot A_c^I$ along with a value $z_c$ that contributes to $\overline{P}$ for the purpose of illustration. By considering how the Minkowski product $E_c^I \cdot A_c^I$ is formed, it is possible to demonstrate that any point on the boundary is generated by a unique pair of points on the boundaries of the two factor intervals. 

To backtrack $z_c$, the first step is to ``invert'' $E_c^I$ to produce a candidate interval $E^I_{c,\text{inv.}}$ of points that can give $z_c$ by multiplication with $A_c^I$
\begin{equation}
    E^I_{c,\text{inv.}} = \left[\frac{|z_c|}{\overline{a_c}}, \frac{|z_c|}{\underline{a_c}} \right] \cdot e^{j(\angle z_c + [-\overline{\Phi_c}, -\underline{\Phi_c}])}.
\end{equation}
The intersection $A^I_c \cap E^I_{c,\text{inv.}}$ meets at the point $A_c$ that will correspond to the maximum coupling strength $|C_{mc}|$, as shown in Fig.~\ref{fig:coupling_backtrack}. Finally, $E_c$ is directly obtained as $E_c = z_c/A_c$. From $E_c$, amplitude and phase error may be found as before. For $A_c$ the phase must necessarily be 
\begin{equation}
\angle C_{mc} = \angle \left\{ A_c - w_c \cdot e^{-j \boldsymbol{k}_s\cdot \boldsymbol{r}_c } \right\} - j \boldsymbol{k}_s\cdot \boldsymbol{r}_m. 
\end{equation}
for all $m \neq c$. For $m = c$ the phase is zero by definition.

\section{\label{sec:stats} Approximate beampattern bounds}

In order to gain a deeper insight into the factors that influence the beampattern bounds and PSLL, an expression for the approximate bounds is derived. The derivation begins with Eq.~\eqref{eq:AP_intervalbeampattern_coupled_simplified}, assuming uniform and symmetric error bounds in amplitude and phase, without directional dependence. To reiterate, this means the maximum phase and amplitude errors are $\pm\delta \Phi$ and $\pm\delta g$, respectively, across all elements. Coupling is also allowed. 

The bounds are derived using the circular interval representation (cIA). First consider the intervals $E_c^I$, which are enclosed with a circular interval, as illustrated in Fig.~\ref{fig:interval_representation}. The radius of the circular interval $R_E$ is the same for all elements $c$, and can be approximated by:
\begin{equation}
    R_{E} = |(1+\delta g) \cdot e^{j\delta\Phi} - 1| \approx \sqrt{\delta \Phi^2 + \delta g^2 }.
\end{equation}
The circular approximation overestimates the true bound, but it is most accurate when $\delta \Phi$ and $\delta g$ produce an annular sector that is well enclosed by a circle (e.g., $\delta \Phi = \pm 3^\circ$ and $\delta g  = \pm 6\%$). Note that $\delta \Phi$ must be expressed in radians in the formula, whereas $\delta g$ is unitless. The geometry being approximated is very similar to the one shown in Fig.~\ref{fig:backtrack_shapes_and_potato}. The deviation from the nominal array response is sought and this deviation is expressed as the sum of the respective radii when using cIA. The particular phase of the array sum is not significant and can be neglected. Therefore, the circle around $E_c^I$, which represents the element sensitivity, can be assumed to have a nominal value of 1.

The circle that represents $E_c^I$ is multiplied with another circle $A_c^I$, which has a radius $R_{Ac}$ (note the dependence on $c$), with the nominal value of the element weighting $w_c$. The product will also have its nominal value on $w_c$. The enclosing circle of the product interval, which forms a Cartesian oval\cite{farouki_minkowski_2001}, has a radius $\rho_c$ 
\begin{equation}
    \rho_c = (1+R_{E})\cdot(w_c + R_{Ac}) - w_c.
\end{equation}

The maximum deviation from the nominal array response is obtained by summing the radii:
\begin{equation}
\begin{split}
 \rho_{B} &= \sum_{c=1}^M \rho_c =  \sum_{c=1}^M (1+R_{E})\cdot(w_c + R_{Ac}) - w_c   \\
  & \gtrapprox R_{E} + \sum_{c=1}^M R_{Ac},
\end{split}
\end{equation}
recalling that $\sum w = 1$. Here the second-order cross-term $R_E\cdot R_{Ac}$ was neglected. To evaluate the sum over $R_{Ac}$, the definitions in Eq.~(\ref{eq:E_and_A_A}) and in the subsequent text can be used, where it can be shown
\begin{equation}
\begin{split}
  \sum_{c=1}^M R_{Ac} & =  \sum_{c=1}^M \sum_{\substack{m=1 \\ m \neq c}}^M \gamma^{|m-c|} \cdot w_m \\
  & = - 1 +  \sum_{m=1}^M  w_m \sum_{c=1}^M\gamma^{|m-c|} \\
  & \lessapprox - 1 + \sum_{m=1}^M  w_m \cdot \left( 1 + \frac{2\gamma}{1-\gamma}  \right) \gtrapprox 2\gamma.
\end{split}
\end{equation}
The following approximations were used; first that the geometric sum runs from $c=-\infty$ to $\infty$, and second the Taylor expansion $\gamma/(1-\gamma) \approx \gamma$.

Finally, the approximate upper bound is obtained as the nominal beam amplitude plus $\rho_B$:
\begin{equation} \label{eq:approximate_bound}
    \overline{P}(\theta) \approx \left( |B_{\mathrm{nom.}}(\theta)| + \sqrt{\delta \Phi^2 + \delta g^2 } + 2\gamma \right)^2.
\end{equation}
The effect of this is that a constant term is added to the nominal amplitude to yield the upper bound. Interestingly, this term does, to a circular IA approximation, not depend on the apodization $\mathbf{w}$. As a result, the worst-case sidelobe level has an asymptotic limit even for infinitely long arrays, which is obtained simply by setting $|B_{\mathrm{nom.}}(\theta)| = 0$ in Eq.~(\ref{eq:approximate_bound}). This is in contrast to the expected beampattern in Eq.~(\ref{eq:expected_P}), which \emph{does} depend on $\mathbf{w}$ through $T_\text{se}$. This result is consistent with the following intuition: In statistical analysis, with independent amplitude and phase errors, the errors add incoherently, so that the average sidelobe level depends on the sum of the variances of the amplitude and phase errors divided by the number of elements. However, for the worst-case analysis, the errors all achieve their maximum allowed values and add coherently, resulting in a bound that does not depend on the number of elements.

\section{\label{sec:results} Numerical experiments}

In this section, we perform numerical experiments in order to illustrate the type of analysis made possible using the theories developed in Secs.~\ref{sec:formulation}, \ref{subsec:direct_verification}, and \ref{sec:stats}. Consider example array B, tabulated in Table~\ref{tbl:arrayB}. For the polygonal method (pIA), the intervals need to be sampled. The sampling is sufficient to make the cumulative representation error well below $-60 \, \mathrm{dB}$. This is calculated as the maximum error in representing the analytic boundaries of the rounded annular sectors (shapes $E_c^I \cdot A_c^I$) and the $M$ number of times these errors are summed. 

\begin{table}[ht]
\vspace{-0.2cm}
\caption{Example array B.}
\vspace{-0.3cm}
\label{tbl:arrayB}
\centering
\begin{ruledtabular}
\begin{tabular}{l r}
Number of elements, $M$ & 31   \\ 
Element pitch & $\lambda/2$ \\
Element diameter $D$ & (omnidirectional) $\approx 0 \lambda$ \\
Array geometry   & Uniform linear array \\
Apodization, $\mathbf{w}$  & $-30 \, \mathrm{dB}$ Chebyshev \\
Steering angle, $\theta_s$ & $-10^\circ$ \\
Maximum amplitude error, $\delta g$ & $\pm 5\%$ \\
Maximum phase error, $\delta \Phi$ & $\pm 5^\circ$ \\
Coupling strength, $\gamma$ & 5\%
\end{tabular}
\end{ruledtabular}
\vspace{-0.2cm}
\end{table}

\begin{figure}[tb]
\includegraphics[width=1\reprintcolumnwidth]{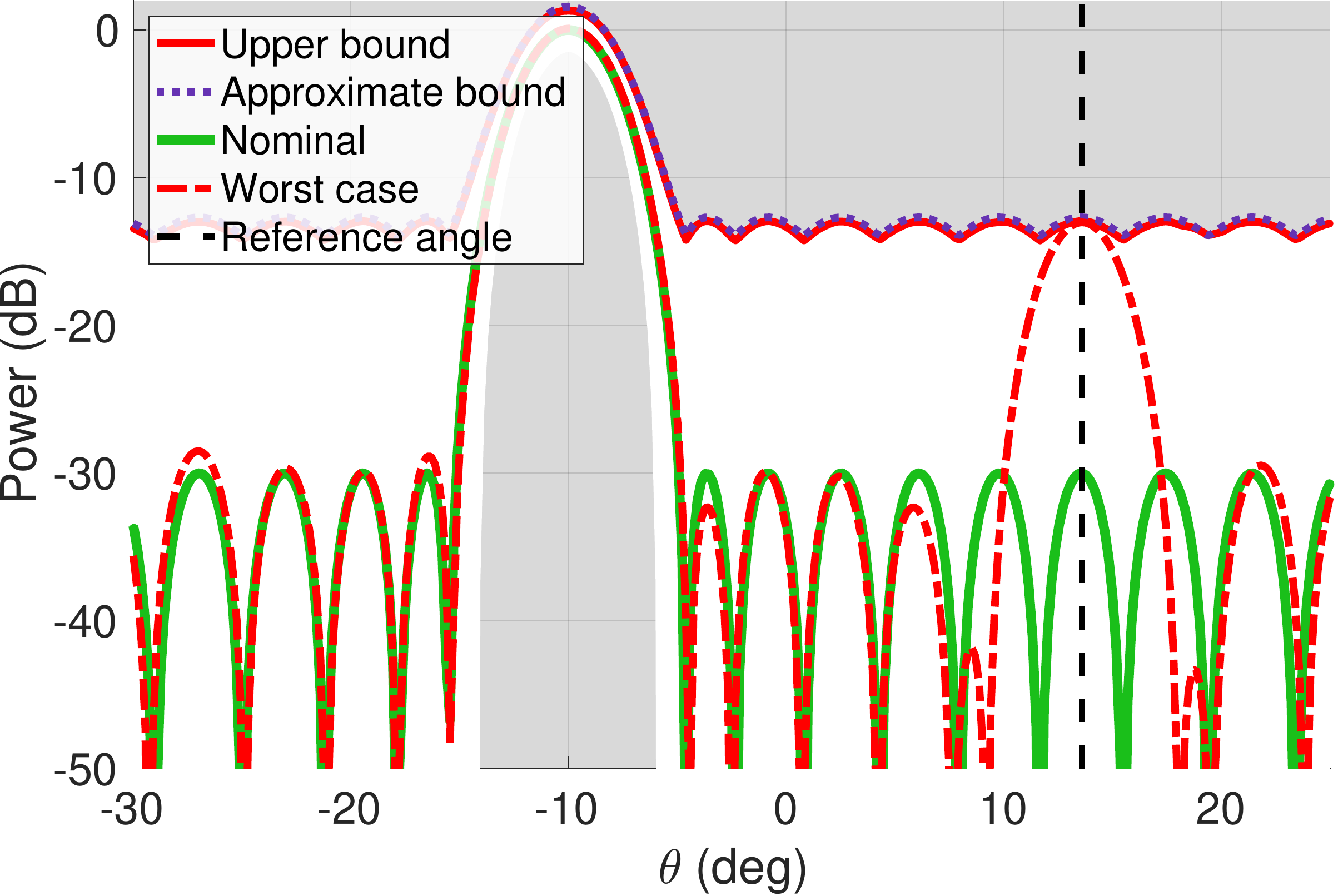}
\caption{\label{fig:exB_bounds}{(Color online) The beampatterns and bounds of example array B, tabulated in Table~\ref{tbl:arrayB}. }}
\end{figure}

The bounds for this problem are shown in Fig.~\ref{fig:exB_bounds}. The lower bound can only be seen close to the mainlobe, and is therefore not mentioned in the legend. The upper bound is relatively uniform, with only minor dips where the beampattern nulls are expected. Additionally, the approximate worst case, calculated with Eq.~(\ref{eq:approximate_bound}), is also illustrated.

\begin{figure*}[tb]
\includegraphics[width=1\textwidth]{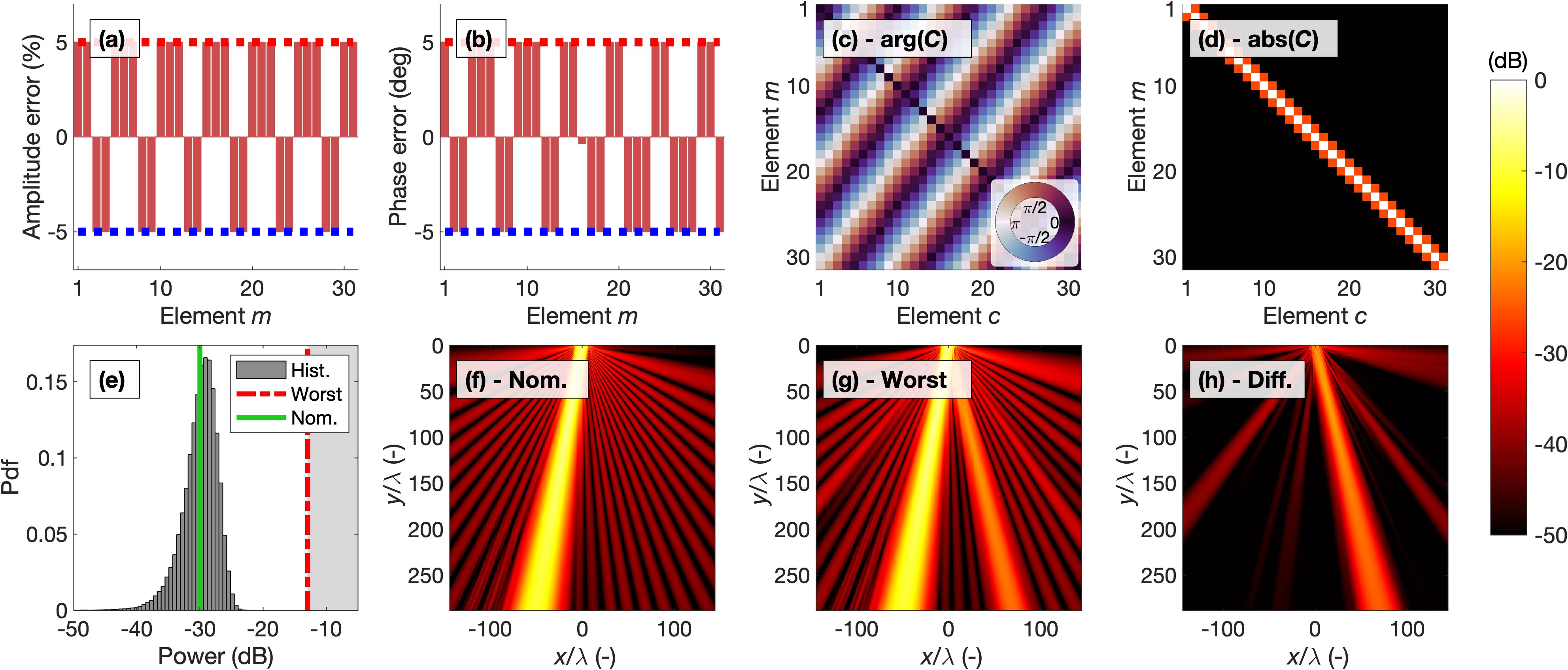} 
\caption{(Color online) Various plots related to the backtracked angle in Fig.~\ref{fig:exB_bounds}. Panel \textbf{(a)} shows the corresponding element amplitude error, \textbf{(b)} the element phase error, \textbf{(c)} the coupling phase error, \textbf{(d)} the coupling magnitude error, \textbf{(e)} the pdf obtained from Monte Carlo simulation of independent  errors within the bounds, \textbf{(f)} the nominal beam, \textbf{(g)} the worst-case beam, and \textbf{(h)} the difference between nominal and worst-case.}
\label{fig:eight_plot}
\end{figure*}

Fig.~\ref{fig:exB_bounds} also shows the worst-case beampattern corresponding to the upper bound at $13.6^\circ$. The backtracking includes both the element errors and the coupling matrix. The backtracked element errors resulting in the particular worst-case sidelobe are shown in Fig.~\ref{fig:eight_plot}\textbf{(a)} and \textbf{(b)}. The backtracked coupling matrix is shown in Fig.~\ref{fig:eight_plot}\textbf{(c)} and \textbf{(d)}. The coupling phase is shown in \textbf{(c)}, and lines are apparent on the anti-diagonals. These lines depend largely on the backtracked angle, and for a certain angle, this matrix will be symmetric. However, in this example, the matrix is only approximately symmetric. Panel \textbf{(d)} shows that only the neighboring elements make a significant magnitude contribution.
The probability density function (pdf) of the array power response is obtained by uniformly sampling the phase and amplitude bounds of element error and coupling, as shown in panel \textbf{(e)}.

To visualize the beam profile, the continuous wave excitation is calculated using the k-Wave function \texttt{acousticFieldPropagator}\citep{treeby_rapid_2018}. The nominal beam is shown in panel \textbf{(f)}, while the worst-case performance is shown in \textbf{(g)}. Plotting the difference beam in panel \textbf{(h)} clearly reveals that the errors align to form a separate beam approximately $-13 \, \mathrm{dB}$ below the mainlobe.

\begin{figure}[tb]
\includegraphics[width=0.9\reprintcolumnwidth]{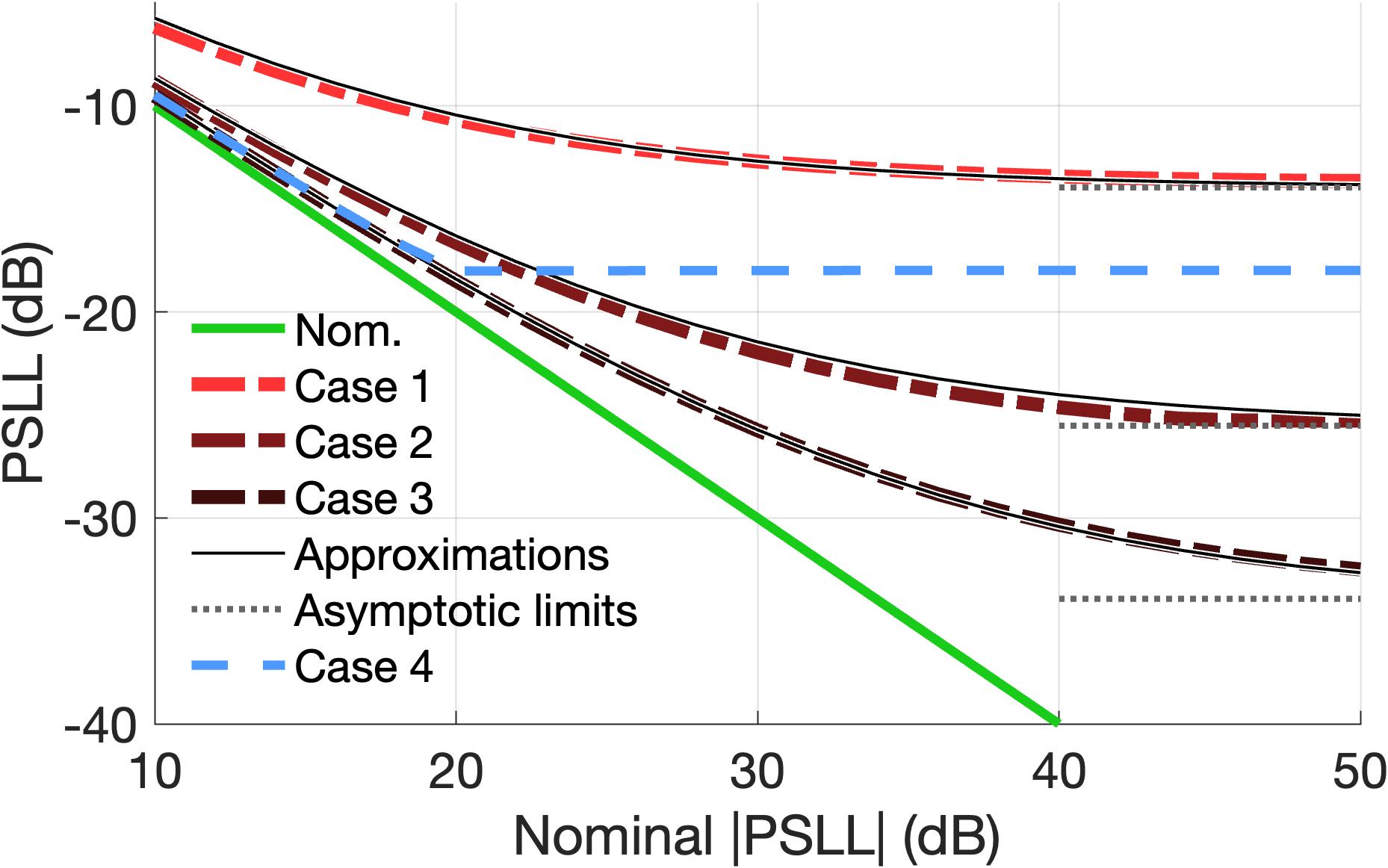}
\caption{(Color online) Worst-case against specified nominal PSLL using Chebyshev apodization. The array geometry given in Table~\ref{tbl:arrayB}. \textbf{Case 1:} $\delta g = 5\%$, $\delta \Phi = 5^\circ$, $\gamma = 5\%$, \textbf{Case 2:} $\delta g = 5\%$, $\delta \Phi = 1^\circ$, \textbf{Case 3:} $\delta g = 1\%$, $\delta \Phi = 1^\circ$,\textbf{ Case 4:} element diameter $D = 0.95 \cdot \lambda/2$, element tilt $\delta \psi = 2^\circ$.}
\label{fig:pssl}
\end{figure}

Finally, it is explored how the PSLL is affected by the nominal sidelobe level specified when using a Chebyshev window. Four different cases are shown in Fig.~\ref{fig:pssl}. Case 1 is the same as discussed earlier (example B). In Case 2, the amplitude and phase error bounds are made more uneven (so the cIA approximation is worse). Case 3 looks at small, but even amplitude and phase error bounds. The analytic value for the asymptotic PSLL as the nominal sidelobes vanish is also plotted, using Eq.~(\ref{eq:approximate_bound}). Case 4 is an outlier; the elements are directive, and a tilt error interval of $2^\circ$ is specified. No approximate bounds are available, but it is evident here that the PSLL decreases until it suddenly flattens out.

\section{\label{sec:discussion} Discussion}

The backtracking technique showed it could recover the worst-case error realization for a PSLL in a given direction. By reapplying the errors it was made evident that the corresponding beampattern reaches the bounds. It is timely to address the relevance of the bounds. The backtracked errors plotted in Fig.~\ref{fig:eight_plot}\textbf{(a-b)} indicate that the errors are usually extreme, which was also seen in Sec.~\ref{subsec:backtrack_simple}. The number of binary (maximum or minimum) error configurations for amplitude and phase, in panel \textbf{(a)} and \textbf{(b)}, is $2^{31\cdot2} \approx 4.6\cdot 10^{18}$. Even if we assume that only one of these configurations represents \emph{the} worst-case error for a particular direction, it be still considered unlikely to encounter a configuration that is the worst-case in \textit{any} direction. Panel \textbf{(e)} illustrates that with independent errors from uniform continuous distributions between the bounds, the sidelobe level clusters around the nominal value, making the bound practically impossible to achieve.

However, while IA may be seen as pessimistic, the statistical method assuming independence represents an optimistic approach. This is because without knowledge of the true bounded error distribution or covariance/dependence between the errors, it becomes challenging to assess the relative probability of values close to the bounds. IA avoids this issue entirely. Additionally, some arrays are composed of a limited number of blocks/modules with common errors within the blocks. Achieving a worst-case block positioning is much more likely than achieving a worst-case element position, and it naturally imposes a periodic structure in the errors. This can be connected with Chapter 3.1.3 in Ref.\citenum{johnson_array_1993}, where a sinusoidal disturbance essentially introduces a dominating sidelobe. It is also important to note that any error pattern resembling, in some sense, the worst-case scenario is still undesirable. Therefore, a practical usage of backtracking can be to identify configurations (such as periodic errors in linear arrays) that are plausible in manufacturing. We argue that for array designers, the worst-case scenario is relevant as it specifies a guaranteed performance all arrays will meet, and that this information is complementary to the expected behaviour.

In order to obtain the most accurate beampattern bounds, the polygonal representation (pIA) was used. This method requires sampling the boundaries of the complex intervals. For the chosen examples, this did not raise any practical issues. However, if the cumulative error has to stay fixed with an increase in the number elements, denser sampling may be needed. This requirement can potentially result in a significant computational cost unless proper methods are in place to address it. A natural technique to mitigate this issue is to remove vertices that are very close to each other (within some tolerance) between each polygon summation. This approach limits a potential exponential growth in the number of vertices, and preliminary tests have shown it to be a promising technique. However, addressing this issue in detail is beyond the scope of this article. 

In the bound formulation with coupling, as shown in Eq.~(\ref{eq:interval_BP}), coupling coefficient reciprocity $C_{ij} = C_{ji}$ would be expected. However, despite concentrated efforts, no solutions has been found to enforce reciprocity due to the dependence problem in IA. On the other hand, as demonstrated by the backtracking results in Fig.~\ref{fig:eight_plot}\textbf{(c)}, there are situations where the coupling matrix $\mathbf{C}$ is \emph{nearly} symmetric, resulting in small overestimation of the bounds, at least for linear arrays when only nearest-neighbor coupling matters. This means that even if coupling reciprocity could be enforced, the worst-case PSLL when using the Chebyshev window would not be significantly different. 

In Fig.~\ref{fig:pssl}, the PSLL for certain error bounds was plotted against the specified nominal PSLL. For cases 1 to 3, the worst-case PSLL eventually deviates significantly from the nominal PSLL. The approximate bound closely follows the accurate bound, but less so in case 2 where the circular approximation assumed is worse. The general agreement indicates reasonable assumptions in the derivation, which may be helped by intermediate approximations that over- and underestimate the true quantities. In any case, choice of apodization and the number of elements can only to a limited extent reduce the worst-case PSLL. 

The lack of directional errors in the approximate formula is not a major concern; for positional errors it can be included for linear arrays (at the cost of a more complicated expression), but eventually directional errors are translated into amplitude and phase anyhow. Case 4 highlights one special feature of directional errors, as directive elements are included in this case. The sudden flattening PSLL is unusual, but closer inspection reveals a peculiar effect: when the incidence angle is $90^\circ$, the worst-case PSLL is obtained by alternate tilting of the elements such that a grating lobe is produced, effectively undersampling the wavefield. This comes from the directivity function in Fig.~\ref{fig:directivity_function}, but it can be argued that the strong effect shown here is a result of the sharp tapering used.


\section{\label{sec:conclusions} Conclusions}

This article presents a comprehensive framework for calculating inclusive beampattern bounds. In addition to tackling arbitrary geometries, a multitude of error intervals can be specified, such as amplitude, phase, position, directivity, and coupling. This flexible technique does not rely on any assumptions about error distributions or correlations.  

The most notable contribution of this study is ``backtracking'', which allows for the direct recovery of the specific configuration of errors and beampatterns that result in the upper or lower bounds. To the best of our knowledge, this is the first time such a direct method has been proposed to verify the beampattern bounds provided by interval arithmetic. Furthermore, backtracking shows which error patterns across the array are detrimental to the peak sidelobe level, allowing array designers to take measures to prevent them in practice.

In addition, this study presents an approximate formula for the bounds of uniformly bounded errors, assuming no directional dependencies. The derived formula specifies and quantifies the factors that influence the worst-case performance of the array. Notably, as opposed to the expected beampattern (a statistical concept), the worst-case beampattern cannot be improved by increasing the number of elements in the array. Additionally, the effect of the apodization window is limited to reducing the nominal beampattern.

The results obtained in this study pertain to the array farfield, but can easily be generalized to the nearfield. In future work, our aim is to analyze modular arrays. The worst-case performance is highly relevant in such systems since fewer degrees of freedom and imposed periodic errors make the worst-case significantly more likely. Additionally, we plan to quantify interval errors in the context of adaptive beamformers. Another open direction for further investigation is to extend the framework to broadband systems, such as medical ultrasound or synthetic aperture sonar. In the latter, various types of periodic array errors are prominent due to the repetition of the same platform in order to synthesize a large array.

\begin{acknowledgments}
The authors thank Roy Edgar Hansen for interesting discussions on the topic and his comments to this work. 
G.~G, T.~I.~L, and A.~A acknowledge funding from the Research Council of Norway project \emph{Element calibration of sonars and echosounders}, project number 317874. J.~E.~K acknowledges internal research funding from InPhase Solutions AS. H.~K.~A. and G.~G. share the first authorship for this article.

\vspace{-10pt}

\end{acknowledgments}
\bibliography{references2.bib} 

\end{document}